\documentclass[11pt,english]{article}
\usepackage{geometry}
\usepackage{verbatim}
\usepackage{textcomp}
\usepackage{amsmath}
\usepackage{amssymb}
\usepackage{esint}

\makeatletter

\@ifundefined{definecolor}
 {\usepackage{color}}{}

\usepackage{babel}

\usepackage{graphicx}%for figure
\makeatother
\usepackage{hyperref}

\newcommand{\be}{\begin{equation}}
\newcommand{\ee}{\end{equation}}
\newcommand{\bea}{\begin{eqnarray}}
\newcommand{\eea}{\end{eqnarray}}
\newcommand{\beas}{\begin{eqnarray*}}
\newcommand{\eeas}{\end{eqnarray*}}

\def\be{\begin{eqnarray}}
\def\ee{\end{eqnarray}}

\def\>{\rangle}
\def\<{\langle}

\def\ba{\begin{align}}
\def\ea{\end{align}}
\def\bas{\begin{align*}}
\def\eas{\end{align*}}

\begin{document}
\begin{titlepage} 

%\begin{flushright}
%{\small{}Preprint number} \\
%{\small{}Preprint number}
%\par\end{flushright}

\begin{center}
\vspace{2mm}

\par\end{center}

\begin{center}
\textbf{\Large{}}\textbf{Plasmonics enabled atomically thin linearly polarized emitter at room temperature}
\par\end{center}

\begin{center}
Bidisha Roy$^{1,2}$, Mäx Blauth$^{1}$, Siddharth Dhomkar$^{3}$, Michael Kaniber$^{1}$, Vinod M. Menon$^{4,5}$ and Jonathan. J. Finley$^{1}$
\\
 
\par\end{center}

\begin{center}
$^{1}$\textsl{\small{}
Walter Schottky Institute, Technical University of Munich
}\textsl{ }\\
\textsl{\small{}Am Coulombwall 4, 87548 Garching bei Munich, Germany}\textsl{ }\\
%\textsl{\small{}Kandi, Sangareddy, Telengana, India 502285}\textsl{ }\\
%\texttt{\textsl{\small{}sroy@iith.ac.in}}~\\
%
\par\end{center}{\small \par}

\begin{center}
$^{2}$\textsl{\small{}
Centre for Nano Science and Engineering
}\textsl{ }\\
\textsl{\small{} Indian Institute of Science, Mathikere, Bengaluru 560012, India}\textsl{ }\\
%\textsl{\small{}Kandi, Sangareddy, Telengana, India 502285}\textsl{ }\\
%\texttt{\textsl{\small{}sroy@iith.ac.in}}~\\
%
\par\end{center}{\small \par}

%\begin{center}
%$^{2}$\textsl{\small{}Racah Inst. of Physics}\\
%\textsl{\small{}Hebrew University of Jerusalem, Jerusalem 91904 Israel}\\
%
%\par\end{center}{\small \par}
%

\begin{center}
$^{3}$\textsl{\small{}Centre for Nanotechnology, University College London}\\
\textsl{\small{}London WC1H OAH, United Kingdom}\\
%\textsl{\small{}University of Bern, Switzerland}\\
%\texttt{\small{} sarkar@itp.unibe.ch }\\
\end{center}
\begin{center}
$^{4}$\textsl{\small{}Department of Physics, City College of New York}\\
\textsl{\small{}City University of New York, 160 Convent Avenue,  New York 10031, USA}\\
%\textsl{\small{}University of Bern, Switzerland}\\
%\texttt{\small{} sarkar@itp.unibe.ch }\\
\end{center}

\begin{center}
$^{5}$\textsl{\small{}Department of Physics, CUNY Graduate Center}\\
\textsl{\small{}365 Fifth Avenue, New York 10016, USA}\\
%\textsl{\small{}University of Bern, Switzerland}\\
%\texttt{\small{} sarkar@itp.unibe.ch }\\
\end{center}

\vskip 1.5 cm
\begin{abstract}
Two-dimensional transition metal di-chalcogenide semiconductors provide unique possibilities to investigate strongly confined excitonic physics and a plasmonic platform integrable to such materials constitutes a hybrid system that can be of interest to enable manipulation of their cumulative optical properties. Here we report tuning of excitonic emission from monolayer WSe$_2$, mechanically exfoliated on top of a periodic two dimensional plasmonic array of elliptical gold (Au) nanodiscs. By exploiting the polarization-dependent nature of plasmonic resonance of the nano plasmonic array (NPA), the photoluminescence (PL) emission from the overlaid monolayer WSe$_2$ could be significantly manipulated. PL is preferentially enhanced at the NPA covered regions of the flake when excited closer to the plasmonic resonant frequencies and previously unpolarized WSe$_2$ PL emission gained $\sim~$20 up to 40 \% degree of linear polarization at room temperature. Obtaining significant spectral overlap between the PL spectrum of WSe$_2$ and the polarization tunable plasmonic resonance of the NPA plays a crucial role in this observation.                                                                                                                                                      The results demonstrate active tunability of optical emission from WSe$_2$ by using an otherwise passive plasmonic environment and opens the possibility of achieving atomically thin linearly polarized emitters at room temperature. In addition to fundamentally interesting physics of such interactions this can be highly desirable for ultrathin orientation sensitive opto-electronic device related applications.

\vspace{4mm}

KEYWORDS: \emph{plasmonics, 2D TMDC, WSe$_2$, photoluminescence, nanoplasmonic array, polarization.} 
\end{abstract}
\end{titlepage}

\setcounter{footnote}{0}
\renewcommand\thefootnote{\mbox{\arabic{footnote}}}

%%%%%%%%%%%%%%%%%%%%%%%%%%%%%%
\section{Introduction}\label{sec:intro}
Investigating optical processes in optically active low dimensional materials in modified photonic environments remain of great interest and importance in the study of light matter interactions. Owing to their remarkable optical, electronic and quantum properties, the field of atomically thin two-dimensional (2-D) transition metal dichalcogenides (TMDCs) has been looked upon for bringing a new paradigm in materials physics of semiconductors showing surge of promising applications in optoelectronics, photonics and valleytronics (ref.\cite{Butler,wangreview,wang2012electronics,schaibley2016valleytronics} and references therein). From few layers to monolayer limit these materials become direct band gap semiconductors \cite{heinz,splendiani}; reduced dimensionality, strong excitonic binding energies ($\sim$ few 100 meV), preferential crystal symmetries, unique combination of spin and valley degrees of freedom, well demonstrated strong light matter coupling \cite{liu2015strong,dufferwiel2015,schneider2018} etc. make them relevant for both fundamental and applied studies. Plasmonics on the other hand, offers to be an easily integrable platform to these materials providing alterable photonic environment to explore and achieve interesting effects such as enhanced quantum efficiency \cite{agarwal2015fano}, strong plasmon-exciton coupling \cite{agarwal2016strong,gonccalves2018plasmonexciton,Carlson}, electrical tunability of such coupling \cite{agarwal2017a,Biswanath}, deep subwavelength guiding of light \cite{blauth2018coupling}, second harmonic generation covering the entire visible range \cite{Ding}, etc. In this context, investigating integrated 2D- TMDCs – plasmonic hybrid system remain interesting for fundamental studies as well as is critical in realizing novel optical devices with applications such as integrated on-chip quantum nanophotonics, improved light sources, detectors and sensors etc. 
%Some key reports in integrating 2D TMDCs to plasmonic platforms include demonstration of modified emission and Fano like resonance from MoS - bow tie array, Purcell enhancement of emission from MoS using silver nano disc arrays, strong exciton-plasmon coupling in MoS coupled with plasmonic lattice, and electrical tuning of such strong coupling, enhanced optical activity of atomically thin MoSe proximal to nanoscale plasmonic slot-waveguides, deterministic coupling of emitters in 2d TMDCs with plasmonic nanocavity arrays, coupling emission from localized defects in 2D TMDC to surface plasmon polaritons and so on. 

In this paper, we investigated a system that consists of a monolayer WSe$_2$ flake, which is intentionally placed on top of a 2D periodic Nano plasmonic  array (NPA) with elliptical gold nano discs. NPAs are ideal for easy integration and possibility of larger area interaction with the 2D material. Additionally, the nature of inter particle coupling of the localized plasmonic dipoles can be exploited from near-field to intermediate to far-field by adjusting the periodicity of plasmonic elements in the NPAs. This may offer an effective way to control the interaction between the excitonic response and the near or far field dipolar-coupled localized plasmonic resonances. In the last years WSe$_2$ has gained much significance lately amongst the other members of the 2D TMDC family, due to its unique valley polarization and coherence properties \cite{urbaszek2014valley,urbaszek2015coherence,hao2016direct}, optically dark ground excitonic state \cite{heinz2015dark,potemski2016darktuning,danovich2017dark}, and the reports of observing single quantum emitters within its defect states \cite{he2015single,potemski2015single,srivastava2015QD}. Here, however, we report an interesting set of observations as seen from the room temperature PL emission of WSe$_2$ in the absence and presence of a simplistically  tunable plasmonic environment. Using polarization dependent reflection spectroscopy, 2D fluorescence mapping, and polarization dependent photoluminescence spectroscopy, we study this influence and achieve an external control over the emission polarization of WSe$_2$ PL by using the polarization tunability of the plasmonic resonance of the NPA pertaining to the geometry. This paves the path to realizable atomically thin linearly polarized emitters at room temperature.

%%%%%%%%%%%%%%%%%%%%%%%%%%%%%%
\section{Results and Discussions}%\label{sec:intro}

%The nature of such interaction was well discussed in ref. Lamprecht.
 
\begin{figure}[ht]
\begin{center}
\includegraphics[totalheight=0.4\textheight, angle=0]{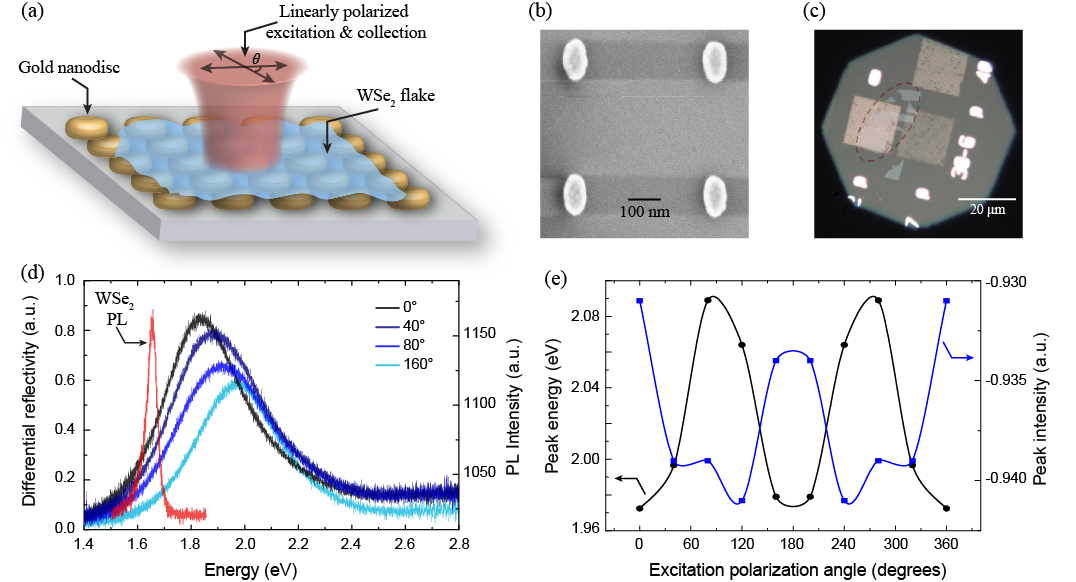}
\end{center}
\caption{(a) A schematic of the sample. (b) SEM image of elliptical gold nano discs fabricated in a two dimensional periodical array (c) Microscopic image of monolayer WSe2 flake mechanically exfoliated on top of the targeted array of elliptical nano discs. (d) Room temperature photoluminescence from bare WSe2 flake overlaid with the broad plasmonic resonance as obtained from the differential reflectivity profile from the selected bare nanoplasmonic array showing that the NPA plasmonic resonance is dependent on the excitation polarization. (e) Dependence of the plasmonic resonance peak energy and intensity as a function of the excitation polarization as obtained from simulations (for details see Supplementary information).
\label{fig:fig1}}
\end{figure}

Elliptical gold (Au) discs (30 nm high, short and long axes radii ranging from 20 to 60 nm) arranged in two dimensional periodic square array lattices (periodicity ranging from 580 to 800 nm) were fabricated using e-beam lithography on 150 µm thick glass substrate. After careful investigation of the fabricated NPAs, monolayer flake of WSe$_2$ was mechanically exfoliated on to a PDMS strip and transferred on top of the NPA of interest comprised of ellipses with 50 nm $\times$ 60 nm short and long axes radii, respectively, arranged in a $\sim~$ 20 µm $\times$ 20 µm 2D periodic array with the periodicity of 600 nm. A schematic of the sample of interest is shown in Fig.1 (a). Details on optimization, fabrication of the nanoplasmonic array and exfoliation of WSe$_2$ flake can be found in the supplementary information. Fig. 1(b) and (c) respectively show a close up SEM image of a suggestive NPA of elliptical discs and an optical microscopic image of the transferred WSe$_2$ flake over the region of interest (highlighted by the red dashed line for eye guidance).

For a single metal nanoparticle the plasmonic resonance strongly depends on details of the the geometric shape and material properties of the particle as well as the surrounding medium. For an ensemble of particles the individual plasmon resonance is additionally influenced by the electromagnetic particle interaction. Depending upon the inter-particle separation between the plasmonic particles, if exceeding the near field coupling distances, typically a far field interaction is mediated by the scattered light fields which are dipolar in nature. For a two dimensional square grating of metal nanoparticles interacting only in the far field regime, particularly strong dipolar interaction arises when the light field corresponding to a particular grating order change from evanescent to radiative in character \cite{meier,lamprecht}. In this work as the 2D NPA is comprised of elliptical gold nano discs with 600 nm separation, the overall plasmonic resonance is a scalar cumulative effect arising from the strong far-field-coupled dipolar interactions between the individual localized plasmonic scatterers. Typical plasmonic response to broad band optical excitation of the NPA is shown in Fig. 1(d). The plasmon resonance is revealed through the measurement of differential reflectivity 
%[R=(R(NPA)-R(substrate))/(R(substrate))], 
$R=\frac{R_{(NPA)}-R_{(substrate)}}{R_{(substrate)}}$
which is the relative change of reflectivity w.r.t that of the substrate. A broadband light source was used for the reflection spectroscopy measurements. After initial adjustment of the excitation power and the polarization, the beam is passed through a beamsplitter and is focused to the sample surface by an achromatic microscope objective (NA = 0.9). To take in account the cumulative scattering from the array a large focused excitation spot (~12 micron diameter) was used by modifying the optics. The reflected light from the sample surface is collected by the same objective, passed through the beamsplitter and guided to a spectrometer where it is spectrally analyzed. Fig 1 (d) shows that the broad plasmonic resonance from the array is dependent on the excitation polarization w.r.t the geometrical axis of the ellipses. In Fig 1(e) the expected behavior of the plasmon resonance as a function of the excitation polarization angle (w.r.t the geometrical axis of the ellipse) is shown as obtained from simulations. It is clear that the plasmon resonance peak intensity and energy go through a periodic oscillation due to the symmetry of the elliptical discs. 
More details on the simulation set up and results can be found in supplementary information. This dependence of the plasmonic resonance on the excitation polarization due to the geometry of the elliptical discs, is one of the key aspects of this work. Figure 1(d) also shows the room temperature PL from bare WSe$_2$ monolayer (PL peak at ~ 1.65 eV). It has been reported that the main exciton emission peak of WSe$_2$ PL at room temperature is from the bound ‘neutral’ exciton with high binding energy;(where the electrons and holes are tightly bound together and are relatively more stable to endure the thermal fluctuations compared to other conventional semiconductors). The PL is overlaid to the plasmonic reflection spectra showing significant spectral overlap between the plasmonic resonance and the excitonic PL at room temperature. 

\begin{figure}[ht]
\begin{center}
\includegraphics[totalheight=0.4\textheight, angle=0]{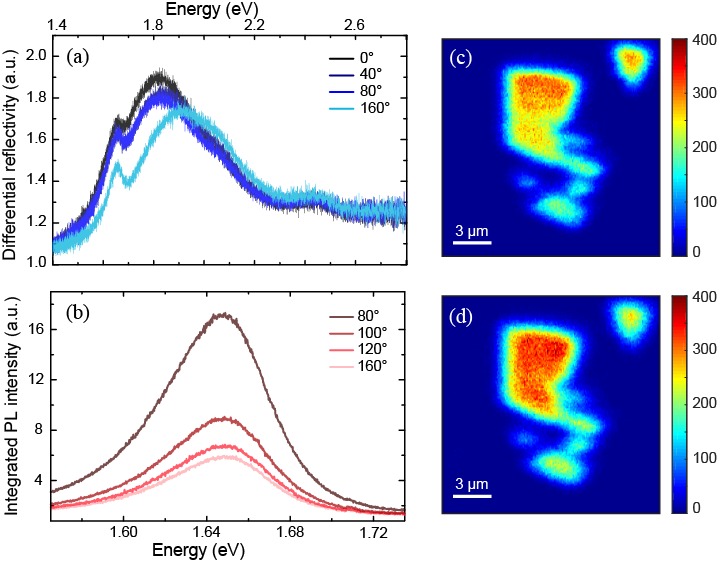}
\end{center}
\caption{(a) Differential reflectivity from nanoplasmonic array covered with monolayer WSe2 revealing the appearance of the absorption edge from the flake. (b) Integrated PL from the WSe$_2$ flaked array excited at the plasmonic resonance showing significant tunability by changing the excitation polarization angle. (c) Confocal fluorescence map of the relevant areas on the sample bare WSe2 flake (right side top triangular region) and that of array overlaying the nanoplasmonic array (larger area) as excited with green (532 nm) laser line and by (d) red (633 nm) laser line, showing preferential enhancement of the PL on the NPA covered region for resonant plasmonic excitation. 
\label{fig:fig2}}
\end{figure}

Differential reflectivity from the WSe$_2$ covered NPA is shown in Fig 2 (a) revealing the appearance of WSe$_2$ absorption edge at the lower energy side in addition to the plasmonic resonance (as seen from the bare plasmonic array). %showed significant overlap of the WSe$_2$ with the plasmonic resonance.  
Fig. 2(b) shows the integrated PL intensity from the WSe$_2$ flaked array excited at the plasmonic resonance (with a 633 nm red laser) for varied angles of linear polarized excitation. This demonstrates that while exciting the WSe$_2$ at the peak of plasmonic resonance a significant tunability of the PL from the overlaying WSe$_2$ flake could be achieved by simply changing the linear polarization of the excitation beam. 

To further investigate into the nature of this interaction, two different excitation sources – a 533 nm green laser (off-resonant w.r.t the plasmon resonance energy) and a 633 nm red laser (resonant with plasmon resonance energy) – were used to obtain a spatial fluorescence map of the WSe$_2$ flake on and off the plasmonic array (we note that both of the excitation are above band gap for WSe$_2$ neutral excitonic transition energy). Fig. 2 (c) and (d) show the fluorescence 2D map with Green and red excitation respectively. The triangular region in upper right side corner is the bare WSe$_2$ flake while the bigger region at the center is the flake covering the NPA. The map clearly shows preferential enhancement of the luminescence intensity from the WSe$_2$ flake overlaying the NPA site as compared to that of the bare flake site only for resonant plasmon excitation (in fig. 2(d)) while for the off plasmonic excitation such an effect is absent (fig. 2(c). We note that this preferential enhancement is not due to any selective increase in effective absorption cross section of WSe$_2$ and hence this preferential enhancement as observed predominantly for the plasmonic resonant excitation is a consequence of the plasmonic environment.

\begin{figure}[ht]
\begin{center}
\includegraphics[totalheight=0.4\textheight, angle=0]{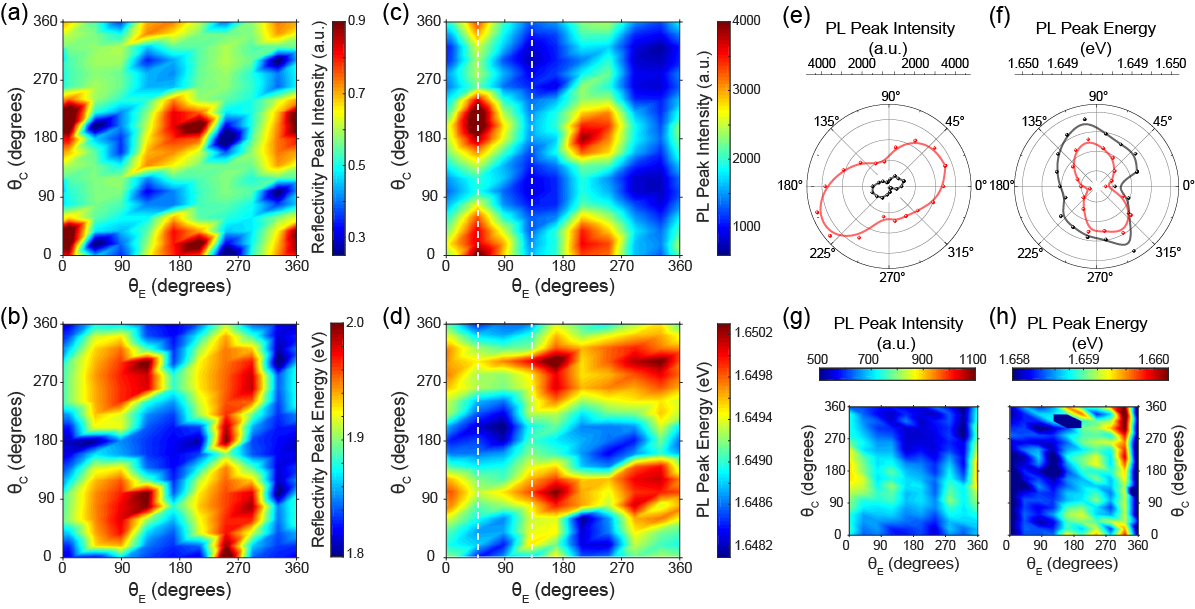}
\end{center}
\caption{Differential Reflectivity (a) peak intensity and (b) peak energy as a function of excitation polarization and collection polarization angles for the bare NPA array (showing the polarization profiles of the plasmonic resonance) (c) Red laser (633 nm) excited PL peak intensity profile for WSe$_2$ flake overlaying the NPA (d) PL Peak energy showing 2 meV spectral tuning for ~200 meV of peak energy shift for the plasmonic resonance. (e) and (f) PL peak intensity and PL peak energy values are as a function of the collection polarization angle for specific excitation polarization angles 45 and 90 degrees respectively (taken from the white dashed line cuts in Fig. 3 (c) and (d)). Photoluminescence (g) peak intensity and (h) peak energy as a function of excitation polarization and collection angles for bare WSe$_2$ flake showing no deterministic linear polarization profile in absence of the NPA. 
\label{fig:fig3}}
\end{figure}
In order to gain further insight on the influence of plasmonic resonant excitation on PL of WSe$_2$, detailed investigations on the peak intensity and peak energy were performed. Fig.3 shows the peak intensity and peak energy of plasmonic resonance and WSe$_2$ PL (for 633 nm i.e. near plasmonic resonant excitation) plotted in a 2D map as a function of the excitation polarization angle ($\theta_e$ on the x-axis) and collection polarization angles ($\theta_c$ on the y-axis). Fig. 3 (a) and (b) demonstrate the geometry dependent scattering response (from differential reflectivity measurement) and polarization profile of the plasmonic resonance from the bare NPA as a function of the excitation and collection polarization angles. High degree of linear polarization is evident, as expected, from the elliptical geometry of the plasmonic particles. It is interesting, however, to note that this polarization property of the palsmonic array directly affects the excitonic PL emission of the WS$e2$ covering the NPA as seen from Fig. 3 (c) and (d). The PL peak intensity and PL peak energy dependence on the excitation polarization angle and collection polarization angle, reveal emergence of significant degree of linear polarization in the emission PL of otherwise unpolarized WSe$_2$ (as also seen in Fig. 3(g)and (h) for bare WSe$_2$ flake where no deterministic linear polarization is observed). This gaining of polarization in the emission PL follows a specific correlation with the polarization response of the plasmonic array which indicates the response is predominantly plasmonics enabled in its nature.  Figs. 3 (e) and (f) show the polar plot for two selected excitation polarization angles (45 and 90 degrees) clearly showing that PL of WSe$_2$ overlaid upon the NPA is linearly polarized in the emission. 

In general, no reports of valley coherence or linear polarized emission for WSe$_2$ has been reported at the room temperature. Nonetheless we measure the degree of polarization upto $\sim~$ 20-40\% for WSe$_2$ on the NPA. We further note that the polarization dependence of the PL does not exactly match with the symmetry of the plasmonic axes (which determines the polarization dependence of the plasmonic resonance itself) while it maintains a slightly shifted maxima-minima correlation with the same. This correlation is dependent on the amount of varying spectral overlap between the plasmonic resonance and excitonic PL spectra. The larger the spectral overlap, higher is the interaction between the plasmonic resonance and excitonic PL and higher is the degree of acquired linear polarization. Essentially, the overall effect can be seen as an additive and dynamic contribution from: 1) the “excitation channel” - where the local field enhancement (due to plasmonic scattering) can cause more efficient excitation of the WSe$_2$ excitons especially at resonant plasmonic conditions and 2) the “emission channel” - where the interplay of the excitation polarization dependent variation of spectral overlap between the plasmonic resonance of NPA and PL of WSe$_2$ facilitates a more dynamic and nontrivial interaction between the two species. We note that  the geometric cross section of the metal filled NPA under WSe$_2$ can be quantified in terms of a “Fill Factor” defined as the following ratio; Geometric Fill Factor = (Area with NPA)/(Area without NPA), which for our system is a merely ~ 2\%. For such a small factor, the contribution from local field enhancement should be insignificant to contribute significantly towards the excitation channel. Thus we conclude that the spectral overlap plays a more crucial role in the effects observed here. The tunability of the amount of spectral overlap between the plasmonic resonance and excitonic emission is the key to the tunability of the acquired non trivial polarization in the emission PL.

%In order to further understand the influence of the plasmonic array on the excitonic PL emission it was necessary to identify if the enhancement is contributed through the excitation channel (due to increase in scattering cross section near the plasmonic resonance) or is it coming from the emission channel (due to the spectral overlap of PL peak energy and plasmon resonance energy). ?? IN this regard….May be another quantification of the amount of spectral overlap and how the PL follows the symmetry of the ellipse axis (and hence the plasmon resonance profile of the NPA).

\section{Conclusion}\label{sec:Conclusion} 
We report nanoplasmonic array enabled modification of PL emission from monolayer WSe$_2$ integrated onto 2D array of elliptical gold nanodiscs.  Using plasmonic resonant excitation the polarization studies revealed that significant degree of linear polarization was imparted to the otherwise unpolarized PL emission from WSe$_2$ at room temperature. This opens up the possibility to obtain atomically thin linearly polarized emitters at room temperature by using array of simple geometrically appropriate plasmonic structures. This can be exploited further towards polarized light sources, flat panel display applications as well as novel alignment sensitivity based sensor applications.

%On excitation with the 633 nm (which is near the plasmonic peak energy) the influence of the NPA on the flake PL is established clearly. 
%following the double extrema nature in both PL peak intensity as well as in PL peak energy (as seen from simulations and experiments in the case of bare NPA).To identify the polarization profile of the degree of polarization of the emission PL, 
%The corresponding plots for PL peak energies are shown in Fig. 3(c) for bare WSe2 and (d) for WSe$_2$ on NPA respectively. 
%Green laser stuff: For both bare WSe$_2$ and WSe$_2$ on NPA the emission polarization profiles are similar in nature and do not exhibit any significant linear polarization. The PL peak energy also remains unaffected for both cases. This confirms that when excited away from the plasmonic resonance, WSe$_2$ PL remains largely unaffected by the effect of the plasmonic enabled polarization tuning. 

%\begin{figure}[ht]
%\begin{center}
%\includegraphics[totalheight=0.3\textheight, angle=0]{test.png}
%\end{center}
%\caption{These are eine buru \cite{Johnson}
%\label{fig:RT}}
%\end{figure}

%\bibliographystyle{unsrt}
%\bibliography{mybib.tex} 

\end{document}